\documentclass[12pt,ukenglish,letterpaper]{article}
\usepackage{amssymb}

\usepackage{graphicx}
\usepackage{amsmath}

\newtheorem{theorem}{Theorem}
\newtheorem{acknowledgement}[theorem]{Acknowledgement}

\input{tcilatex}
\begin{document}

\title{Protein folding in high-dimensional spaces: hypergutters and the role
of non-native interactions}
\author{T.C.B. McLeish \\
Department of Physics and Astronomy and Astbury Centre for Molecular Biology%
\\
University of Leeds, Leeds LS2 9JT, UK}
\maketitle

\begin{abstract}
\setlength{\baselineskip}{0.26in}We explore the consequences of very high
dimensionality in the dynamical landscape of protein folding. \
Consideration of both typical range of stabilising interactions, and folding
rates themselves, leads to a model of the energy hypersurface that is
characterised by the structure of diffusive ``hypergutters'' as well as the
familiar ``funnels''. \ Several general predictions result: (1) intermediate
subspaces of configurations will always be visited; (2) specific but \textit{%
non-native} interactions are important in stabilising these low-dimensional
diffusive searches on the folding pathway; \ (3) sequential barriers will
commonly be found, even in ``two-state''proteins; (4) very early times will
show charactreristic departures from single-exponential kinetics; (5)
contributions of non-native interactions to $\Phi $-values are calculable,
and may be significant. \ The example of a three-helix bundle is treated in
more detail as an illustration. \ The model also shows that high-dimensional
structures provide conceptual relations between the ``folding funnel'',
``diffusion-collision'', ``nucleation-condensation'' and ``topomer search''
models of protein folding. \ It suggests that kinetic strategies for fast
folding may be encoded rather generally in non-native, rather than native
interactions. \ The predictions are related to very recent findings in
experiment and simulation.
\end{abstract}

\setlength{\baselineskip}{0.26in}

\section{\protect\bigskip Introduction}

The current conceptual map of protein folding kinetics is dominated by the
coexistence of several apparently distinct approaches. \ They may be
categorised loosely into ``energy landscape'' (Bryngelson et al., 1995;
Onuchik, 1995), ``diffusion-collision'' (Karplus and Weaver, 1976, 1994),
``nucleation-condensation''(Fersht, 2003) and ``topomer search'' (Makarov
and Plaxco, 2002) models. Each of these has its own way of visualising how
the collapse of a random coil to a native globule can ever be accomplished
in observable time scales, a problem pointed out long ago (Karplus, 1997). \
Each has advantages and drawbacks, but it is not clear whether each applies
to a restricted subset of real cases, or whether all might have something to
say about the folding of any one protein.

The ``folding-funnel'' picture of the energy landscape has the advantage of
visualising both guided folding and the emergence of on-pathway and
off-pathway intermediate states (Dinner et al. 2000). \ Yet it is hard to
escape from the deceptive simplicity of low-dimensional projections of
folding funnels that appear necessarily in all graphical portrayals of it. \
In practice of course, the dimensionality of the folding space is enormous.
\ Even small ($\sim 100$ residue) proteins have\ a configurational space
dimensionality of several hundred (think of the bond angles along the
polypeptide main chain alone). \ In such high-dimensional spaces,
qualitatively new features may arise, such as energetically-flat domains
that nonetheless are extremely difficult to escape from and so behave as
kinetic traps. \ A second feature is the potential for high cooperativity of
structure in several simultaneous dimensions. \ This corresponds to the
existence of narrow gullies in the hypersurface that are hard to find. \ In
more biochemical language these structures might be exemplified by
cooperative secondary structure formation alongside native or near-native
distant contacts in $\alpha $-helix bundle proteins (Myers and Oas, 2001),
or simultaneous folding and anion binding (Henkels at al 2001).

The ``diffusion-collision'' approach, on the other hand, is supported by
strong experimental evidence that folding rates are controlled by the rate
of diffusion of pieces of the open coil in their search for favourable
contacts, rather than a driven collapse along some continuous energy surface
(Jacob et al, 1999; Plaxco and BAker, 1998; Goldberg and Baldwin, 1995). \
Pre-formed units of secondary structure diffuse hydrodynamically and merge.
\ Larger proteins may do this in an increasingly hierarchical way. \ The
importance of diffusive searches is unsurprising, since under biological
conditions, all candidates for energetic interactions, including
electrostatics, are \textit{locally} screened to a few angstroms: much
smaller than the dimensions over which sections of protein must move to find
their native configurations. \ Put another way, the vast majority of the
space covered by the energy landscape must actually be \textit{flat }(on a
scale of $k_{B}T$) rather than funneled.\ \ Simple versions of these models
have indeed been able to account rather well for folding rates as a function
of secondary structure formation (Myers and Oas, 1999, 2001). \ However, it
is not clear how applicable this approach is to cases in which secondary
structure forms \textit{within} a collapsed globule or cooperatively with
it. \ 

An attempt to articulate a range of scenarios in which partial ordering of
secondary and tertiary structures mutually enhance a favourable folding
pathway has been presented under the label of ``nucleation-condensation''
(Dagett and Fersht, 2003). \ Originally conceived as a kinetic theory in
which a nucleus of native structure corresponds to the transistion-state for
folding, the picture now also encompasses the hierarchical folding routes of
the diffusion-collision model.

A challenge faced by all these models is that the most successful search for
inherent features of tertiary structure that correlate with folding rates
has found that the topological measure of ``contact order'' is far more
closely related than, for example, molecular weight itself in the case of
``two-state'' folders (Plaxco et al 2000). \ Rationalisation of this
observation has given rise to a third view of the critical pathway of
protein folding, the ``topomer search'' model (Plaxco and Gross 2001). \ The
rate determining step is not the rapid formation of local secondary
structure, nor the diffusion of subdomains \textit{per se}, but the
organisation of large pieces of secondary structure into the same
topological configuration as the native state, which is thereafter is able
to form rapidly. \ This suggests a partition of the folding space into
``rapid'' dimensions representing the local formation of secondary
structure, and ``slow'' dimensions representing the topomer search. However,
a quantitative relation between the topomer search space and contact order
is still unclear, since no native contacts are actually required to form at
a purely topologically-defined transition state at all (although many are to
be expected from the patial ordering at the secondary level at least). \
Furthermore, information on the effect on folding rate of replacing specific
residues \textit{via} mutation or ``$\Phi $-value'' analysis (Fersht, 2000)
needs to be taken together with correlations of contact order.

These four approaches have one important aspect in common: they all
effectively reduce the dimensionality of the search-space by assumption,
rather than in a derived way. \ This is both natural and necessary, since
data from kinetic experiments do just the same, but there is a danger in
overlooking aspects of folding that rely essentially on the presence of many
degrees of freedom. \ Our aim in this work is to take a fresh look at the
issue, embracing many simplifications but on this occasion \textit{not} that
of a low dimension of configurational search space. \ We find in the next
section that quite general conclusions may be drawn about the topology of
this search space if the dimensionality is kept high. \ Some general
predictions follow which we work out in more detail in the case of
three-helix proteins. \ The approach will additionally allow us to see how
the existing apparently-distinct paradigms for protein folding are related,
and suggest places to look for the information content of the ``kinetic
code'' within proteins that encodes the folding search path, as distinct\
from the native structure itself.

\section{High-Dimensional Diffusive Subspaces and Searches}

We start with a very simple and abstract model for protein folding, but one
that explicitely retains a very large number of degrees of freedom. \ The
total search space is modelled as the interior of a hypersphere of dimension 
$d$ and radius $R$, and the native (target) state as a small sphere of
radius $R_{N}$ at the origin of the space. The entire configuration of the
protein corresponds to a single-point particle executing a random walk in
the hypersphere. \ \ The ratio of $R$ to $R_{N\,}\,$describes the typical
localisation on folding in the values of a degree of motional freedom. \ \
If the degree of freedom is spatial the appropriate scales are the size of a
molten globule and the radius of gyration of a denatured protein. \ If it is
angular, then they are the angle of libration of a bond fluctuating in one
local minimum as a fraction of $2\pi $. \ In either case the appopriate
order of magnitude estimate is which is $\left( \frac{R}{R_{N}}\right)
\simeq 10$. \ Bicout and Szabo (Bicout and Szabo 2000) introduced this very
general framework for discussing flat and funneled landscapes, but then
restricted themselves to three-dimensional spaces, a simplification that we
shall try to avoid. \ To explore the timescales of the search for the target
space (on which the diffuser will ``stick'') we write down the
time-dependent diffusion equation for a particle, restricting ourselves to
the case of a flat potential at first. \ The most convenient function to use
is the probability density $P\left( r,t\right) $, that the system is a
radial distance $r$ from the centre of the hypersphere at time $t$, which
obeys:

\begin{equation}
\frac{\partial P\left( r,t\right) }{\partial t}=D\frac{1}{r^{d-1}}\frac{%
\partial }{\partial r}r^{d-1}\frac{\partial P\left( r,t\right) }{\partial r}
\label{highddiff}
\end{equation}%
supplemented by the absorbing boundary condition $P(R_{N},t)=0$, signifying
the stability of the native state. \ The timescale for the search steps is
set by the effective diffusion constant $D$. \ The mean passage time from
the unfolded ensemble to the native state can be calculated by introducing a
uniform current $J$ of diffusers (representing a population of folding
proteins) on the boundary of the hypersphere at $r=R$, as the other boundary
condition, and finding the consequent steady state solution to (\ref%
{highddiff}). \ The mean time to pass from $R$ to $R_{N}$ over the ensemble
of systems is then just the total number of diffusers at steady state
normalised by the current, leading to 
\begin{equation}
\tau _{f}=\frac{1}{d\left( d-2\right) }\left( \frac{R}{R_{N}}\right)
^{\left( d-2\right) }\frac{R^{2}}{D}  \label{search time}
\end{equation}%
This expression indicates how very much \textit{qualitatively} longer the
mean search time is in \textit{high} dimensions ($d>2)$, than the
low-dimensional estimation of the characteristic time $\tau \simeq R^{2}/D$,
which replaces (\ref{search time}) in $d=1,$ and $2$. \ This fundamental
time is scaled up by the denatured system size (measured in units of the
target size $R_{N}$) to the power of the number of effective dimensions
greater than 2. \ An analysis of the eigenmode structure of the problem
indicates why this is so: for large $d$ nearly all the diffusers exist in
the lowest eigenfunction of the diffusion operator, that is in turn
localised to the exponentially large surface of the hyperspherical search
space. \ Single-exponential kinetics are also a general property of such
high-$d$ search spaces.

The central result of (\ref{search time}) depends on two key physical
assumptions: (1) the dimensionality of the space is of realistic values for
protein folding - of the order of a hundred or more, and (2) the stability
of the folded state is governed by local interactions in the native state
only. \ With these assumptions alone, the model of high dimensional
diffusion we have described is inevitable, and the timescales unreasonably
long. \ The exponentially large search times arise transparently from the
factor $\left( \frac{R}{R_{N}}\right) ^{\left( d-2\right) }$ in equation (%
\ref{search time}). \ This is of course a restatement of Levinthal's paradox
(Karplus, 1997), but a helpful one, in that the two necessary assumptions
for the paradox to arise are clearly seen. \ The first just gives the large
dimensionality of hypersphere, the second the flat diffusive landscape. \ 

Put this way, there are two ways of circumventing the problem. \ One may
drop the assumption of local forces and allow the protein to ``fall''
towards the single native state down a ``funnel'' created by forces whose
range permeate the entire volume. \ As we have remarked above, however,
candidates for such long range forces do not present themselves. \ Without
recourse to a continuous funnel-shaped landscape, there is only one other
possibility: \textit{all diffusive searches take place in low dimensional
subspaces }of the full configurational space.

To see how this works, we suppose at first that the $d\sim O(10^{2})$
dimensions of the full folding space are now arranged sequentially so that
diffusive searches in one dimension at a time allow the protein to find
``gateways'' into the next subspace (we will see how this may arise
naturally in a physical way below). \ For simplicity we assume that the
kinetics of each diffusive search is single exponential with characteristic
time $\tau $. \ Since the diffusion is always maintained in some
low-dimensional subspace of the full folding space, $\tau \sim R^{2}/D$ for
each subspace, so that $\tau _{fold}\simeq d^{2}R^{2}/D$ rather than the
exponentially larger $\left( R^{2}/R_{N}^{2}\right) ^{d}$. \ This clearly
reduces the folding time enormously, signifying that only a tiny fraction of
possible states is visited in the search (Dinner et al., 2000).

How has such a remarkable reduction\ in folding time been achieved without
the use of a ``funnel'' energy landscape? \ Of course energetic interactions
have been implied, but these have not been of the spatially-extended
``funnel'' type. \ Instead they have served just to keep the diffusive
search within the smaller $\left( d-1\right) $ or $(d-2)$ dimensional space,
once the first sub-dimensional search is over, then within a $\left(
d-3\right) $ or $(d-4)$ dimensional space after the subsequent successful
``adsorption'' into the still-smaller subspace, and so on. \ So, when the
high-dimensionality of the search space is retained, the energy landscape
looks less like a funnel, and more like a series of high-dimensional \textit{%
gutters} (figure 1). \ The diffusing particle (representing of course the
random search of the protein through its available conformations) does not
have to search simultaneously through both the dimensions of the figure. \
Instead, it \ exploits the lower energy state of the entire $(d-n-1)^{th}$
dimensional subspace to reach it \textit{via} a \textit{one}-dimensional
diffusion in the $(d-n)^{th}$ dimension, which it performs first. \ By
partitioning the configurational space in this way, and by providing an
attractive ``gutter'', relying on \textit{local} forces alone, to connect
one diffusive subspace to the next, all the advantageous consequences of a
funnel landscape may be acquired without the requirement of long-range
potentials. \ Of course, if the high-dimensional structure is projected into
a 1 or 2 dimensional diagram then the many discrete steps of potential
energy that arise from the sequence of ``hypergutters'' appear artificially
close, and serve to create a funnel-like projected energy landscape. \ The
disadvantage of the projection is that the subtle origin of the directed
search is obscured. \ In detail the folding energy landscape will look more
like a series of low-dimensional \textit{terraces} (inset to figure 1)
nested within the full high-dimensional search space.

How big do the attractive potentials creating the gutters need to be, and
what physical interactions might be enlisted to provide them? \ Their scale
is familiar: these potential steps are just the energies required to
counterbalance the entropy-loss associated with reduction of the
configuration space by one dimension, or degree of freedom. \ The associated
translational space reduces from the order of $R$ to the order of $R_{N}$ on
restriction to the gutter subspace. \ Completely reversible folding along
the route connecting the gutters is produced by rendering the free energy
change on entering the gutter zero. \ This is in turn the case if the
binding energies to the gutters are of the order of the entropic free energy
gain on making such a restriction to a degree of freedom:

\begin{equation}
\Delta U_{gutter}\simeq k_{B}T\ln \left( \frac{R}{R_{N}}\right)
\label{gutterpot}
\end{equation}%
To quantify $\Delta U_{gutter}$ therefore needs just an estimate of the
order of magnitude of the ratio $\left( \frac{R}{R_{N}}\right) $, the
dimensionless ratio of the sizes of space enjoyed by a degree of freedom in
and out of a restricting gutter. \ As discussed above, a realistic order or
magnitude estimate is $\left( \frac{R}{R_{N}}\right) \simeq 10$, giving a
value for $\Delta U_{gutter}$ of the order of a few (2-4) $k_{B}T$ (or of
order 4-8 kJmol$^{-1}$) for realistic proteins. \ The energy scale of a few $%
k_{B}T$ is highly suggestive: we note that the relatively weak, non-native
like interactions between residues are candidates for these
gutter-stabilising interactions, and that it is not necessary to invoke the
strength of native contacts during the diffusive search. \ This is good
news, since there is no guarantee that significant native interactions will
form during at least the early phases of search, if at all, and experimental
evidence of strong ``co-operativity'' is to the contrary (Flanagan et al.,
1992). \ \ Of course we do not assume that the energy-entropy balance is
exact at each step - indeed it is the mismatches in this picture that give
rise to roughness in the landscape, but matching within a few $k_{B}T$ is
necessary in most dimensional reductions to avoind unrealistically long
folding times.

Furthermore, the evolutionary tailoring of non-native interactions provides
additional ``design-space'' within which a pathway to the folded state may
be coded, but without compromising the stability of the final, native state.
\ For proteins containing $N$ residues, there are of order $N^{2}$
non-native interactions that may be encountered during a diffusive search,
but only of order $N$ interactions that define the native state. \ A second
consequence of this high-dimensional viewpoint is therefore the general
expectation of tuned\textit{\ }but\textit{\ non-native, interactions}
between sections of partially structured chain that stabilise\textit{\
intermediate search spaces} (which may or may not be identified as
intermediate states, depending on their occupancy lifetime). \ We need to
articulate carefully what is meant in this context by ``non-native'', for
this term is sometimes used to refer to indiscriminate interactions. \ In
that sense, the role of non-native interactions in determining the type and
rate of folding pathways is not a new idea (Zhou and Karplus, 1999). \ But
such previous studies have not introduced any specificity, or evolutionary
refinement, into the non-native interactions, and find, significantly, that
increasing the strength of such indiscriminate interactions actually slows
folding. \ Our suggestion is that a \textit{discriminating design} of key
non-native interactions may significantly speed the search for the native
state. \ It is also likely that a significant proportion of such tailored
non-native interactions that we envisge guiding the search will be
increasingly near-native as the search proceeds. \ This will be the more
likely as secondary structure forms, as we shall see by the example of a
three-helix buundle below.

Gutter-like landscapes have appeared in the literature, and are sometimes
apparent even in the 2-dimensional representations of projected folding
surfaces. \ Reference (Karplus and Weaver, 1994), for example, shows a fast
folding route of hen lysozyme in which the early formation of $\beta $-sheet
structure permits the final approach to the native state to proceed in a
subspace of reduced dimension. \ In this case the gutter-like structure
survives a projection onto just two dimensions of folding space. \ In this
case the mutual diffusion of the helical and beta-sheet portions of the
protein is the dynamical process responsible for the gutter-like feature on
the reduced folding surface. \ This example serves also to indicate an
important qualification - some dimensions clearly \textit{do} possess
funnel-like landscapes even without a projection onto low dimensional
spaces. \ Those involved with the formation of a local $\alpha $-helix or $%
\beta -$turn structures, for example, create subspaces that have real
funnel-like features, directed towards the point in the subspace
representing the formation of the complete local secondary structure. \
However, higher-dimensional hypergutters must already have been visited at
higher levels in the regions of locally $\alpha $ and $\beta $ secondary
structure. \ We now take a much simpler fold as an example.

\section{ An example: three-helix proteins}

\ A clean example of a ``hypergutter'' structure is furnished by the
well-studied triple-helix proteins such as the B-domain of staphylococcal
protein A (BdpA) (Myers and Oas, 2001) (and see figure 2). \ In this case,
the division of the folding landscape is clearly suggested by the formation
of the helices (fast ``funneled'' or ``zipper'' dimensions (Fiebig and Dill,
1993), and by the diffusive search of the helical domains for their native
juxtaposition. \ Note that we do not require the helix formation to be
complete before the diffusive search begins - indeed the formation of native
or non-specific contacts and secondary structure stability will in general
be highly cooperative (Fersht, 2000). \ All that is required is that the
zipper dimensions are explored at much faster timescales than the diffusive
dimensions. A very simple model has been successful in describing the
kinetics of this protein (Myers and Oas, 2001), using the physical
abstraction normal to ``diffusion-collision'' models of the real protein as
spherical domains executing a spatial search. \ Such models have recently
been extended to a family of three-helix bundle proteins (Islam et al.,
2002). However, in the light of our expectation that fast-folding proteins
find their native state \textit{via }a sequence of stabilised subspaces, the
diffusive degrees of freedom of a three-helix bundle might be more
accurately represented by angular coordinates defined at the two turns
connecting the three helical sections. \ In fact the diffusive space of
internal angles thus defined is exactly three dimensional: between helix 1
and 2 only one angle needs be specified, while between helices 2 and 3 we
need two more. \ This construction is illustrated as in figure 2.\ The three
angular diffusive degrees of freedom are labelled $X_{i}$ with $i=1,2,3$. \
Since the diffusive coordinates are angles, they exhibit periodicity, and
the search space is itself a periodic 3-d lattice. \ In practice the
continuously varying angular co-ordinate may model a more discrete set of
more or less favourable packings (Chothia et al., 1981), but the
coarse-grained structure of the search space will be the same. In the figure
we illustrate periodicity in the dimension $X_{2}$ only. \ The region of
configuration space in which the first two helices are both in contact with
the third is shaded, and the native state is represented by the periodic
lattice of small spheres. \ If the shaded ``helical contact'' region is
enhanced by a weak attraction (it becomes a ``gutter'' for diffusion in the $%
X_{2}$ coordinate), then the search for the native state will typically
proceed by diffusion in the one-dimensional manifold of $X_{2}$ (without
contact between helices), followed by diffusion in the two-dimensional
manifold of $X_{1}$ and $X_{3}$ (now with helices 1 and 3 in contact). As
calculated in the last section, the non-native binding potential of the
third helix to the gutter sub manifold needs to be of the order of $3k_{B}T$%
. \ Providing that the gutter is as attractive as this, then the predicted
mean search time (including prefactors and a weak logarithmic term) for the
native state is%
\begin{equation}
\tau _{1/2}=\left( \frac{R^{2}}{D}\right) \left[ \frac{16}{\pi ^{2}}+\frac{1%
}{2}\left( \ln \frac{R}{R_{N}}-1\right) \right]   \label{tau12}
\end{equation}%
rather than the much longer time for the full 3-d search without the gutter
subspace of%
\begin{equation}
\tau _{3}=\left( \frac{R^{2}}{D}\right) \left( \frac{R}{R_{N}}\right) \left( 
\frac{4}{3\pi ^{2}}\right)   \label{tau3}
\end{equation}%
Examples of experimental evidence for staged diffusive searches in simple
proteins has also been observed in the case of cytochrome C, lysozyme (Bai,
2000), and in the B1 domain of protein G (Park et al., 1999).

The 3-helix example illustrates our general conclusion that searches within
diffusional subspaces in protein folding may be accelerated by\textit{\
local, but not necessarily native, interactions} between sections of
partially structured chain. \ In the context of the three-helix protein the
necessary non-specific interactions are those that keep the third helix in
contact with the other two. \ This permits the final diffusive search for
the native state to take place in a 2-dimensional space, rather than the
full 3-dimensional search\ configuration space of the diffusive degrees of
freedom. \ \ Remarkably, just this conclusion was reached very recently by
experiments on the helical immunity protein Im7 (Capaldi et al., 2001), in
which an on-pathway intermediate state was shown by careful mutation studies
to be stabilised by non-native interactions between two of the helices. \ An
additional example of tuned non-native interactions guiding a folding
pathway occurs in the rather larger Phage 22 Tailspike protein (Robinson and
King, 1997), where a non-native disulphide bond controls the folding search.
We remark that in both these cases, the stabilised hypergutter provides an
arena in which ``diffusion-collision'' calculations can operate within a
molten globule, \ so constituting a significant generalisation of that model
to non-spatial degrees of freedom (Zhou and Karplus, 1999).

\section{Predictions of the Hypergutter model}

We have identified two general predictions of this high-dimensional view of
folding: (1) the sequential diffusive exploration of low-dimensional
subspaces favoured by fast folding and (2) the stabilisation of these
subspaces by discriminate but non-native (or near-native) interactions,
without recourse to long-range guiding forces. \ But it has other things to
say concerning common experimental measures of even the deceptively simple
``two-state'' folders. \ We derive here three further consequences: (3)
early-time structure in kinetics, (4) temperature and denaturant dependence
and the free-energy structure of the folding pathway, (5) non-native
contributions to $\Phi $-values.

\subsection{Relaxation functions in folding kinetics}

\ We first take a very simple case: if the non-native gutter-stabilising
interactions are perfectly balanced with the entropy changes at each stage
of the dimension reduction, then the free-energy profile is itself flat, and
the $d$ diffusive dimensions form an effective one dimensional path along
which the folding takes place. \ This is not, of course, to suggest that the
path is unique, since: (i) a large fraction of each sub-dimension may be
explored, (ii) the path is at each stage reversible and (iii) the
non-diffusive ``zipper'' dimensions describing the local folding of
secondary structure are perpetually exploring their own configurational
space rapidly and cooperatively with the slow dimensions. \ Nonetheless,
casting the high-dimensional problem into this form shows that a na\"{\i}ve
``reaction co-ordinate'' picture can actually emerge from the concatenation
of the sequentially-stabilised hypergutters.

An effective 1-dimensional coordinate, $X$ arises from such concatenation of
the gutter dimensions of a very high dimensional space, whose initial
condition (for a quenching experiment) will favour the high entropy of the
early dimensions: every initial state is completely disordered, and the
resulting one-dimensional diffusion equation will be supplemented by the
approximate initial condition $p\left( X,0\right) =2\delta \left( x\right) $%
. \ If the native state is representes by a sink for diffusers at $X=1$, it
is straightforward to calculate the fraction of unfolded proteins after a
quench as:%
\begin{equation}
\sum_{n=0}^{\infty }\frac{4\left( -1\right) ^{n}}{\pi \left( 2n+1\right) }%
\exp \left( -\left( 2n+1\right) ^{2}t\right) 
\end{equation}%
which we plot as a solid line in figure 3. \ For most of its trajectory,
this function mimics a single exponential, but with an effective \textit{%
delay} from the moment of quench. \ This arises from the time it takes for
the higher subspaces to be filled - at first the native configuration is
``screened'' by virtue of being buried in a cloud of states of low entropy.
\ The apparent delay would be noticed only in experiments able to capture
the very fastest kinetics after a quench.

\subsection{Temperature dependence and the folding pathway}

It is very common to represent diagrams of the folding and unfolding rates $%
k_{f}$ of proteins plotted against the concentration of denaturant, the
so-called ``chevron plot''. \ A more challenging experiment is represented
by the Eyring plot ($\ln k_{f}$ with $1/T$). \ These plots do not typically
show the simple linear form characteristic of chemical reactions with a
transition state free energy that is itself independent of temperature. \ In
the case of protein folding reactions they are generally negatively curved,
and may contain discontinuities of gradient (Oliveberg et al., 1995). \ A
common interpretation is to claim that the local gradient of the Eyring plot
gives the activation enthalpy at that temperature, and therefore that
curvature implies a change in the enthalpy of the transition state. \
Possible causes suggested have included melting of the differently-sized
hydration shells in the unfolded and transition states.

Such a curvature in Eyring plots is, however, a natural consequence of the
hypergutter model. \ Low-dimensional diffusive subspaces of progressively
lower entropy are stabilised by increasingly negative non-specific binding
energies. \ On average the energies (enthalpies) of the subspaces towards
the native state must have large negative values, since the folded native
state has such a low entropy compared to the fully denatured state. \ The
enthalpies $\Delta H_{n}$ of the $n^{th}$ diffusive subspace must attempt to
counterbalance their increasingly large (negative) entropies $T\Delta S_{n}$%
. \ 

Without any further information on the implicit optimisation of these sets
of energies and entropies, but with only one overall constraint that the
total free energy to fold $\Delta G_{D\rightarrow N}$ be a fixed value (at
some temperature, zero at the folding temperature $T_{f}$), we can define,
as in the last section, a free energy pathway as one-dimensional random walk
through the diffusive subspaces (see figure 4). The transition state is the
diffusive subspace with the highest free energy. \ From the figure, it is
clear that this corresponds to the greatest positive excursion of the random
free energy trajectory. \ It is also clear that, at the folding temperature $%
T_{f}$ (figure 4(b)) this tends to occur midway through the walk, at the
point least controlled by the boundary conditions at the endpoints. \ Well
into the folding regime of $T<T_{f}$, however, this maximum excursion is
much more likely to be near the unfolded state, $D$. \ 

Calculations of the mean excursions of a random walk of $Q$ steps of an
average energy-difference $\epsilon $ and constrained to a total (negative)
free energy change $\Delta G_{N}$, can be parameterised by the dimensionless
quantity $x\equiv \left( \frac{\Delta G_{N}}{\epsilon }\right) Q^{1/2}$ (see
appendix \ref{randomTSapp}) . \ In terms of $x$, the expectation value of
the transition state free energy is the surprisingly universal form%
\begin{equation}
\frac{\left\langle \Delta G^{\ddagger }\right\rangle }{\epsilon Q^{1/2}}=%
\sqrt{\frac{1}{4}\left( 1-\sqrt{\frac{x}{1+x}}\right) \left( 1+\sqrt{\frac{x%
}{1+x}}\right) }{}-x\left( 1-\sqrt{\frac{x}{1+x}}\right)   \label{TSmax}
\end{equation}%
This form may be recast as an Eyring plot using $k_{f}\sim \exp \left( \frac{%
\Delta G^{\ddagger }}{k_{B}T}\right) $, together with the assumption of a
linear dependence on temperature for the depth of the native state energy
near the folding temperature itself, so that $x=\beta \left( 1-\frac{T}{T_{f}%
}\right) $, with $\beta $ a dimensionless number. \ The dimensionless Eyring
plot for the parameter $\beta =1$ is given in figure 5. \ The curvature of
the plot comes from the mean shift of the transition state to more denatured
diffusive subspaces along the folding pathway. \ Alternatively, it can be
interpreted as a particular prediction for the Hammond shift of the
transition state for high-dimensional protein folding with only the minimal
requirements of \ overall entropy and enthalpy balance at each entry down
the sequence of gutters. \ Qualitatively, the effect is close that seen in
several cases, and with a comparable magnitude (Oliveberg et al., 1995). \ 

There are, of course, several \textit{caveats} attached to such a general
calculation. \ Clearly this crudest form cannot pick up specific and
discontinuously large shifts of the transition state, which in small
proteins will often dominate particular cases (see, for example, the two
regimes of temperature for which continuously-curved Eyring plots hold in
the case of barnase (Oliveberg et al., 1995)). \ Nor does it anticipate
specifically evolved favourable departures from the random imbalances of
entropy and energy assumed here, which certainly arise and roughen the
landscape, nor does is account for specific behaviour from hydration shells.
\ Nonetheless, we can see that the qualitative features of
temperature-dependent folding do arise without any special assumptions of
these kinds. \ Furthermore, it suggests a rather general structure for the
free-energy along a folding pathway, in which successive fluctuations in
entropy and energy create a sequence of intermediate states. \ This type of
structure has been investigated theoretically (Wagner and Kiefhaber, 1999)
and evidence for its rather general emergence has arisen experimentally very
recently (Pappenberger et al., 2000; Sanchez and Kiefhaber, 2003).

\subsection{Phi-Values: non-native contributions}

As a final general prediction, and as an example of the specific
calculations possible with the model, we examine the important question of $%
\Phi $-value analysis and its interpretation. \ When the mutation of a
residue gives rise to a value of $\Phi $ close to $1$, it means that the
change in folding rate arising from the mutation is consistent with
comparable changes in the transition state and native state energies. \ This
is usually interpreted to mean that the residue in question enjoys the
majority of its native contacts at the transition state (Fersht, 2000). \ \
However, this model suggests another physical source of positive values for $%
\Phi $, since it identifies \textit{non-native} interactions as potentially
crucial in establishing folding rates. \ For if a residue contributes 
\textit{via} non-native interactions to the stable hypergutter \textit{that
concludes the dominant (longest) diffusive search}, then mutations to that
residue will affect the folding rate, even though it does not necessarily
possess any native contacts at the transition state. \ In the hypergutter
model, the ``transition state'' is, by definition, the subspace following
that which takes the longest time to search - the rate-determining step. \ 

To make this more precise, we return to the case of the three-helix bundle
and calculate the dependence of the total folding rate on the non-native
potential that stabilises the 2-dimensional ``gutter'' of the final search.
\ Defining a ``fugacity'' $\Delta =\sigma e^{\varepsilon /kT}$, where $%
\varepsilon $ is the stabilising energy of the 2-d gutter and $\sigma
=\left( \frac{R_{N}}{R}\right) $, the measure of the relative sizes of the
two spaces, we expect that as $\Delta $ is increased (by increasing $%
\varepsilon $) we take the system from the the slower 3-dimensional search
to the accelerated ``1+2'' dimensional search. \ By adding the currents of
diffusers that find the native state from the 2 and 3 dimensional spaces
separately, we find an approximate cross-over formula for the folding rate $%
k_{f}$ (ignoring weak logarithmic factors):%
\begin{equation}
k_{f}=\left( \frac{\Delta }{1+\Delta }\right) k_{1+2}+\left( \frac{1}{%
1+\Delta }\right) k_{3}  \label{rate crossover}
\end{equation}%
where $k_{1+2}=\tau _{1+2}^{-1}$ and the slow rate $k_{3}=\tau
_{3}^{-1}\approx \sigma k_{2}$. \ The expression (\ref{rate crossover}) also
contains, by implication, a prediction of the contribution of the \textit{%
non-native} interactions to the $\Phi $-values of the residues that
contribute to it. For, a mutation of any residue will change its
contribution to the gutter potential, so%
\begin{equation}
\Phi _{g}=\frac{1}{n_{g}}\frac{\partial \ln k}{\partial \varepsilon }\approx 
\frac{\Delta }{n\left( 1+\Delta \right) \left( \sigma +\Delta \right) }
\label{gutter phis}
\end{equation}%
where $n_{g}$ is the number of residues that share the burden of providing
the non-native gutter potential $\varepsilon $. \ We have also assumed in
the derivation of (\ref{gutter phis}) that $\varepsilon $ is also the scale
of a \textit{single} residue's contribution to the stability of the native
state - but other reasonable assumptions will only introduce an order-$1$
prefactor. \ The functional dependence of $\Phi _{g}$ on the fugacity $%
\Delta $ is actually a rather weakly varying function once the gutter is
large enough to produce a reasonable fraction of the maximum acceleration of
the folding rate, and is close to the value $0.6/n_{g}$ (it possesses a
broad maximum of this value when $\Delta \simeq \sqrt{\sigma }$, see figure
6, which shows how both folding rate and $\Phi _{g}$ depend on the gutter
potential). \ This non-native contribution to $\Phi $ will naturally be weak
in the two limits of vanishing gutter-potential (when all searches are
high-dimensional) and very high gutter potential (when they are always
low-dimensional). \ \ 

Again, a rather general result emerges that may be compared with rate
measurements on selectively-mutated systems. \ For the three-helix bundles,
contributions to the stabilising potential that encourages the terminal
helices to diffuse in contact with each other will arise typically from one
residue per helical turn, so that $n_{g}\precsim 10$ (by counting about two
residues per turn on the contact face of a 5-turn helix). \ Since the total
predicted non-native contribution to $\Phi $ is or order $1$ (from the
dimensionless function of (\ref{gutter phis}) plotted in figure 6), this
means in turn that mutating these residues might generally give non-native
contributions to their apparent individual $\Phi $-values of order 0.1 . \
The inset to figure 6 displays the expected pattern of such enhanced $\Phi $%
-values against residue index. \ Remarkably, this is precisely what is seen,
again in very recent experiments, on the immunity family of helical proteins
Im7 and Im9: the member of the family with an on-pathway intermediate (Im7)
also exhibits increased $\Phi $-values in the appropriate region of the
helices 1, 2 and 4 (helix 3 only forms co-operatively with the native state)
by just this amount, relative to the protein without the intermediate, Im9
(Friel et al., 2003).

The magnitude of incremental contributions to $\Phi $ from the gutter
potentials is restricted to these low values only in very simple topologies
such as the three-helix bundle. \ When key stabilising interactions succeed
in reducing the dimensionality of the search space more drastically, much
higher values can result (from differentiation of the higher-dimensional
analogues of (\ref{rate crossover})). \ In more complex spaces of mutual
diffusion of helices and $\beta $-turns, values greater than $1$ are not
unexpected. \ This approach suggests a natural interpretation of $\Phi $%
-values greater than one, such as recorded in acylphosphatase (Chiti et al.,
1999), but which do not bear an interpretation in terms of native structure
(Fersht, 1999). \ 

The interpretation of non-classical $\Phi $-values outlined here is
closely-related to a recent suggestion arising from some simple lattice
Go-type simulations (Ozkan et al., 2001). \ The simulation also found that
kinetic properties are more closely connected with $\Phi $ than local
``degree of nativeness''. \ It shares with the present treatment the
essential departure from a one-dimensional projection of a transition state,
and an identification of the number of permissable pathways, or transition
entropy, in controlling the rate of folding.

\section{Relationship to other models}

The model of high-dimensional diffusive hypergutters is not incompatible
with the frameworks or results of the other models discussed in the
introduction, but rather serves to show how the apparently alternative
models are related. \ Each emerges from the high-dimensional hypergutter
picture when a \textit{different} projection into a low-dimensional space is
applied. \ 

When the flat, diffusional, freedom are projected away onto a reaction
co-ordinate pair such as $R_{g}$ and $\phi _{native}$, then a folding funnel
appears, and does so without the presence of any long-range forces. \ The
difference is that, on close examination, the funnel is discrete, or
terraced, rather than continuous. \ Furthermore, it appears when the
interactions generating coil-collapse are projected along an ordinate of
sequential sub-spaces, rather than along a spatial co-ordinate. \ But when
there are many sequential subspaces, an apparently continuous folding funnel
appears with all of the features of intermediate states, multiple pathways 
\textit{etc.} ascribed to it (Brygelson et al., 1995; Dinner et al., 2000)
arising in a natural way. \ Another example of this projection is found in
the master-equation approach (Zwanzig, 1995), in which the smoothly funneled
high dimensional energy landscape implies some shaping of the non-native
contacts of the underlying model.

When the projection is \textit{orthogonal} to one of the later diffusional
subspaces, on the other hand, then the same system will appear to map onto a
diffusion-collision model. \ In this case the projection concentrates on the
diffusive degrees of freedom one by one, rather than projecting them away
into a funnel. \ The advantage of the hypergutter approach, however, is that
it identifies diffusive subspaces in cases where the standard
diffusion-collision model does not. \ The case of diffusion in mutual
angular space of helical bundles discussed above is an example, since this
occurs within a globule, rather than in the collisional formation of a
globule. \ It also recognises intermediate cases in which diffusional
searches occur simultaneously in high and low-dimensional spaces, such as a
partially-stabilised 2-d gutter in the three-helix case, and provides a
structure for introducing tailored, rather than indiscriminate, non-native
interactions.\ \ The interesting and unexpected prediction of non-native and
positive $\Phi $-values emerges in just this case.

Of particular interest is the relationship of the hypergutter picture to the
topomer search model. \ This is because the rate-determining diffusive
searches will in general be completed only when a topological, as well as a
spatial, constraint in the final native state is satisfied for the first
time. \ It is also to be expected since this model also seeks to understand
folding rates without recourse to their dependence on native interactions. \
Again, the three helix bundle serves as a model example- restriction from
the 3-d helical angular space to the faster 2-d space with helices 1 and 3
in contact occurs when the topological orientation of the helices is
satisfied. \ Similar conclusions would emerge if the slow-searches were
between a helix and a $\beta $-sheet, or between two or more $\beta $-turns.
\ Retaining all the diffusional degrees of freedom leads to a close
relationship between the contact order $Q_{D}$ of the topomer search model,
and the number of diffusional dimensions $d$ of the space of hypergutters in
the rate-determining diffusive search. \ For we can identify the exponent $%
Q_{D}$ in the rate expression in the topomer search model (Makarov and
Plaxco, 2003)%
\begin{equation}
k_{f}\sim \gamma \left\langle K\right\rangle ^{Q_{D}}  \label{topomer rate}
\end{equation}%
with that in the diffusive-search result (\ref{search time}) above to find,
formally at the level of the exponent that $Q_{D}=\left( d-2\right) $ where
here $d$ is the dimension of the largest diffusive search. \ We note,
however, that departures from the correlation of folding time with contact
order might be expected when non-native interactions are tuned to speed up
folding in the way we have outlined. \ This is because such a strategy can
reduce the effective dimension of the search without changing the topology
of the final state. \ Strong outlying behaviour in the correlation of
folding time with contact order may be connected with the potential
variability in the efficiency of hypergutter stability portrayed in figure 6.

\section{Discussion and Conclusions}

We have discussed a conceptual approach to the protein landscape problem
that attempts to remain faithful to the high dimensionality of the system. \
\ Rather than invoking a continuous energy landscape with long-range forces
giving rise to a funneled landscape, we use rather general considerations to
point to a high dimensional structure of ``hypergutters''. \ These
structures describe the search for the native state as a sequence of
relatively low-dimensional diffusive subspaces. \ Only spatially-local
interactions are required to direct the folding towards the native state in
reasonable times. \ As a by-product, this procedure also draws together into
a single picture the apparently divergent views of the folding funnel,
collision-diffusion, nucleation-condensation and topomer search models. \ \
The rate-determining ``gutter'' dimensions lie orthogonal to other
``zipper'' dimensions describing the local formation of secondary structure,
that \textit{are} characterised by a continuous folding funnel. \ 

Looked at another way, our structure is a more detailed examination of the
sort of dynamic processes that must be occurring within the ``molten
globule'' phase of protein folding. \ The formation of the globule itself
from the denatured state corresponds in this picture to the first
hypergutter in a series. \ It is clearer in experiment than the subsequent
dimensional reductions because it is the only one that makes significant
changes to the radius of gyration of the protein.

We might note that such a pattern of non-specific binding in diffusive
searches is a common motif in biology, appearing for example in the search
of DNA-binding repressors for their operons (Winter et al., 1981). \ In this
case, the slow search for a specific binding site in $d=3$ is substituted
for a much more rapid diffusive search in $d=1$ (along the DNA) by
non-specific binding of the repressor proteins. \ In this process too, there
is strong evidence that the non-specific interactions are themselves subtly
coded to further speed the search for the binding target.

Several general predictions follow. \ The first is that special tuning of
non-native interactions may contribute significantly to rapid folding; they
stabilise the hypergutter-potentials that keep diffusive search dimensions
under control. \ In the case of helical proteins, candidates for the
structure of the gutters are non-specific contacts of the helices, and the
angular, rather than translational, degrees of freedom describing their
mutual configurations. \ In proteins with more complex structures, other
candidates suggest themselves, such as the orientation of helices with
respect to $\beta $-sheets with which they are in contact in $\alpha /\beta $
proteins, and the relative orientations of $\beta $-turns and
partially-folded $\beta $-sheets in all-$\beta $ elements. \ Very recently,
the role of non-native interactions in stabilising an on-pathway
intermediate, together with a diffusion-collision kinetic route, has been
experimentally verified in the case of the immunity protein Im7 (Capaldi et
al., 2002). \ This precisely exemplifies the general mechanism we have
suggested, with the additional feature that one of the hypergutters has
become so stabilised (and therefore so populated) that it attracts the label
of ``intermediate state''. \ Other examples of non-natively stabilised
folding pathways are emerging (Robinson and King 1997). \ Perhaps the most
remarkable example is the determination by fast kinetic experiments that $%
\beta $-lactoglobulin employs an transient helical motif that is entirely
non-native (Park et al., 1999). \ By stabilising a $\beta $-turn that
otherwise relies on highly non-local, and late-forming, structure for its
stability, this temporary helix reduces the dimension of the search space
for the non-local contacts. \ Of course, it is not impossible to achieve the
dimensional reduction we have outlined by using fully native, rather than
non-native interactions. \ Such proteins would present a highly bimodal
distribution of $\Phi $-values, clustering closely to $0$ and $1$. \ A
candidate would be acylphosphatase (Paci et al., 2002) in which the
``transition state'' strongly constraints the environment of just three
residues and the immediate neighbours. \ In the hypergutter picture, the
nine-dimensional search space of these critical regions separately is
reduced to three sequential three-dimensional searches by the long lifetimes
of the native regions when the remainder of the protein is disordered. \ One
possible advantage of using broadly distributed non-native interactions,
rather than a few local native ones, to stabilise hypergutters, is that the
intermediate states are thereby more tightly confined. \ This is turn may
assist in suppressing the pathway to aggregated states, or amyloid formation
(Dobson, 2002). \ This suggestion has recently been made from observations
of competing folding pathways in a $\beta $-sandwich protein

As both the sensitivity and time-resolution of kinetic experiments
increases, finer details of the intermediate diffusional subspaces in this
and other proteins should become equally transparent. \ Another recent,
theoretical, contribution has pointed out that fine-structure of a few $%
k_{B}T$ within the transition state on a reaction pathway can accelerate
folding (Wagner and Kiefhaber, 1999). \ As an example of the type of
fine-structure predicted, the model contains a natural explanation of the
generic curvature seen in Eyring plots of the temperature dependence of
folding rates. \ The pseudo-random differences in the enthalpy and entropy
of the transitions from one diffusive subspace to the next result in the
transition state energy becoming typically more and more negative as the
quench depth increases. \ This Hammond-like behaviour will contribute at
least in part to the experimental signals, and suggest the gathering of a
wider data sets of this type. Very recent findings suggest that a structure
very similar to this predicted one is indeed rather general (Sanchez and
Kiefhaber, 2003).

Features of the time-dependent folding curves as functions of temperature or
denaturant also follow from the model, including the possibility of an
apparent delay before single exponential kinetics set in. \ It is also
possible that ``kinetic traps'' arise not just from low-energy intermediate
states, but from intermediate diffusive subspaces of higher dimension than
2, for which the control of dimensionality has been incomplete. \ This is
significant for the topomer search model: we might expect to find departures
from the folding time/contact order correlation when, in spite of sharing
the same topology of fold, one protein in a pair has an important
diffusional subspace stabilised while the other does not. \ Alternatively,
our picture suggests ways of increasing folding times greatly by selective
mutations that retain topology and stability of the native state, but
destabilise one or more of the on-pathway diffusional subspaces so that
intermediate searches are required in $d>3$.

The model provides an alternative interpretation of the results of protein
engineering analysis, and suggests that not all contributions to measured $%
\Phi $-values at the transition state arise from native-like interactions. \
It suggests interpretations for $\Phi $-values of order 0.1-0.2, but also
indicates that contributions to larger values (including the non-classical
range $\Phi >1$) may arise from non-native interactions with that residue
that serve to restrict the folding space. \ More detailed predictions of
non-native contributions to $\Phi $ for the family of bacterial immunity
proteins and their mutants are in accord with very recent experiments.

Finally it suggests that the ``kinetic code'' that informs the search for
the native state may be found in evolved selection of some of the non-native
interactions. \ The large number of these, of order $N$ times greater than
the number of native interactions, make them a likely candidate for
information storage, as well as their natural propensity for kinetic control.

The framework and specific examples discussed here suggest useful
coarse-grained models of other families of proteins that may be simulated
very efficiently, or even approached analytically, as we have done with the
3-helix bundles. \ Strong experimental evidence is currently emerging that
supports all of the main predictions of the approach; other experimental
tests of the more surprising conclusions are awaited.

\appendix

\section{\label{randomTSapp}Diffusive subspaces and random energy excursions}

We consider the free-energy as a directed one-dimensional walk $W$ in energy
space of $Q$ steps each of mean (free) energy $\epsilon $. \ We first
consider the case of return to the origin, which corresponds to the folding
temperature $T_{f}$ at which $\Delta G_{D\rightarrow N}=0$. \ To find the
typical excursion of the walk we may map the problem onto a one-dimensional
polymer (so that the free-energy walk itself has a free energy $F\left(
W\right) $). \ At a coordinate $z=n/Q$, corresponding to the $n^{th}$
diffusional subspace of the $Q$ total, this free energy of the walk is
composed of the segment before $x$ and the segment after. \ If $\Delta G(z)$
is the free energy at coordinate $z$, the free energy of the whole walk,
integrating over all other possible energies for the individual subspaces is
the sum of the two ``entropic elastic'' contributions from these pieces:%
\begin{equation}
F\left( W\right) =\frac{\left( \Delta G(z)\right) ^{2}}{\epsilon ^{2}Qz}+%
\frac{\left( \Delta G(z)\right) ^{2}}{\epsilon ^{2}Q\left( 1-z\right) }
\end{equation}%
Since the probability for energy excursions $\Delta G(x)$ is proportional to 
$e^{-F(W)/k_{B}T}$, the distribution of $\Delta G(x)$ at each point is
Gaussian, with mean%
\begin{equation}
\left\langle \Delta G_{\max }(z)\right\rangle =\epsilon Q^{1/2}\left(
z-z^{2}\right) ^{1/2}
\end{equation}%
which peaks, of course, at $x=1/2$.

Now taking the directed random walk, the expected maximum value of the free
energy at $x$ is just the random excursion calculated above superimposed on
a steady drift towards the native state $x=1$ at the native free energy. \
Taking out constants with dimensions we write:%
\begin{equation}
\frac{\left\langle \Delta G_{\max }(z)\right\rangle }{\epsilon Q^{1/2}}%
=\left( z-z^{2}\right) ^{1/2}-xz
\end{equation}%
where the parameter $x=\frac{\left| \Delta G_{D\rightarrow N}\right| }{%
\epsilon }Q^{1/2}$ is a measure of the quench depth. \ It is a simple matter
to maximise this function over $z$, and then to find the maximum value. \
The result is equation \ref{TSmax}.

\bigskip

{\Large References}

Bai., Y. 2000. Kinetic evidence of an on-pathway intermediate in the folding
of lysozyme. Protein Sci. 9:194-196

Bicout, D.J. and A. Szabo. 2000. Entropic barriers, transition states,
funnels, and exponential protein folding kinetics: a simple model. Protein
Sci. 9: 452-465

Bryngelson, J.D., J.N. Onuchic, N.D.Socci, and P.G. Wolynes. 1995. Funnels,
pathways and the energy landscape of protein folding. Proc. Struc. Func.
Gen. 21:167-195

Capaldi, A.P. , C. Kleanthous, and S.E. Radford. 2002. Im7 folding
mechanism: misfolding on a path to the native state. Nature Struct. Biol.
9:209-216

Chiti, F., N. Taddei, P.M. White, M. Bucciantini, F. Magherini, M. Stefani,
and C.M. Dobson. 1999. Mutational analysis of acylphosphatase suggests the
importance of topology and contact order in protein folding. 1999. Nature
Struct. Biol. 6:1005-1009

Chothia, C., M. Levitt, and D. Richardson. 1981. Helix to Helix Packing in
Proteins. 1981. J. Mol. Biol. 145; 215-250

Dagett, V., and A.R.Fersht. 2003. Is there a unifying mechanism for protein
folding? TIBS 28:18-25

Dinner, A.R., A. \v{S}ali, L.J. Smith, C.M. Dobson, M. Karplus,. 2000.
Understanding protein folding via free-energy surfaces from theory and
experiment. TIBS 25:331-339

Dobson, C.M. 2002. Protein misfolding, evolution and disease, TIBS,
24:329-333

Fersht A.R., 1999. Structure and Mechanism in Protein Science: A Guide to
Enzyme Catalysis and Protein Folding, W.H. Freeman, New York

Ferscht, A. R., Transition state structure as a unifying basis in protein
folding mechanisms: contact order, chain topology, stability and the
extended nucleus mechanism. 2000. Proc. Natl. Acad. Sci. USA 97:1525-1529

Fiebig, K.M. and K.A. Dill. 1993. Protein core assembly processes. J. Chem.
Phys. 98:3475-3487

Flanagan, J.M., M. Katoka, D. Shortle, and D.M. Engelman. 1992. Truncated
staphylococcal nuclease is compact but disordered. Proc. Natl. Acad. Sci.
USA 89:748-752

Friel, C.T., A.P. Capaldi, and S.E. Radford. 2003. Structural Analysis of
the Rate-limiting Transition States in the Folding of Im7 and Im9:
Similarities and Differences in the Folding of Homologous Proteins. J. Mol.
Biol. 326:293-305

Goldberg, J.M. and R.L. Baldwin. 1995. Diffusional barrier crossing in a
two-state protein folding reaction. Proc. Natl. Acad. Sci. USA 96:2019-2024

Henkels, C.H., J.C. Kurz, C.A.Fierke, and T.G. Oas. 2001. Linked folding and
anion binding of the bacillus subtilis ribonuclease P protein. Biochemistry
40:2777-2789

Islam, S.A., Karplus, M., and D.L. Weaver, D.L. 2002. Application of the
diffusion-collision model to the folding of three-helix bundle proteins. J.
Mol. Biol. 318;199-215

Jacob, M., M. Geeves, G. Holtermann, and F.X. Schmidt. 1999. Diffusional
barrier crossing in a two-state protein folding reaction. Nature Struct.
Bio. 6: 923-926

Karplus, M. The Levinthal paradox: yesterday and today. 1997. Fold Des.
2:S69-S76

Karplus, M., and D.L. Weaver. 1976. Protein folding dynamics. Nature, 260:
404-406

Karplus, M., and D.L. Weaver. 1994. The diffusion-collision model and
experimental data. Protein Sci. 3:650-668

Kuwata, K., M.C.R. Sashtry, H. Cheng, M. Hoshino, C.A. Batt, Y. Goto, and H.
Roder. 2001. Structural and early kinetic characterisation of early folding
events in $\beta$-lactoglobulin, Nature Struct. Biol. 8:151-155

Makarov, D.E., and K.W. Plaxco. 2003. The topomer search model: a
quantitative, first-principles description of two-state protein folding
kinetics, Protein Sci. 12:17-26

Myers, J.K. and T.G. Oas. 1999. Reinterpretation of GCN4-p1 folding
kinetics: partial helix formation precedes dimerisation in coiled coil
folding. J. Mol. Biol. 289: 205-209

Myers, J.K., and T.G. Oas. 2001. Preorganised secondary structure as an
important determinant of fast protein folding. Nature Struct. Bio. 8: 552-558

Oliveberg,Y.-J. Tan, and A.R. Fersht. 1995. Negative activation enthalpies
in the kinetics of protein folding. Proc. Natl. Acad. Sci. USA, 92, 8926-8929

Onuchic, J.N., P.G. Wolynes, Z.LuthySchulten, and N.D. Socci. 1995. Toward
and outline of the topography of a realistic protein folding funnel. Proc.
Natl. Acad. Sci. USA, 92:3626-3630

Ozkan , S.B., I. Bahar , and K.A. Dill. 1999. Transition states and the
meaning of -values in protein folding kinetics. Nat. Struct. Biol. 8:765-769

Paci, E., M. Vendruscolo, C.M. Dobson, and M. Karplus. 2002. Determination
of a transition state at atomic resolution from protein engineering data. J.
Mol. Biol., 324:151-163

Pappenberger , G., C. Saudan, M. Becker, A.E. Merbach, and T. Kiefhaber.
2000. Denaturant-induced movement of the transition state of protein folding
revealed by high-pressure stopped-flow measurements. Proc. Natl. Acad. Sci.
USA 97:17-22 (2000)

Park, S.-H., M.C.R. Shastry, and H. Roder, 1999. Folding dynamics of the
B1domain of protein G explored by ultrarapid mixing. Nature Struct. Biol.
6:943-947

Plaxco, K.W. and D. Baker. 1998. Diffusional barrier crossing in a two-state
protein folding reaction. Proc. Natl. Acad. Sci. USA 95: 13592-13596

Plaxco, K.W., and M. Gross. 2001. Unfolded, yes, but random? Never! Nat.
Struct. Biol. 8:659-670

Plaxco, K.W., K.T.Simons, I. Ruczinski, and Baker, D. 2000. Contact order,
transition state placement and the refolding rates of single domain
proteins. Biochemistry 39: 11177-11183

Robinson, A.S., and King. 1997. Disulfide-Bonded Intermediate on the Folding
and Assembly Pathway of a Non-Disulfide Bonded Protein. Nat. Struct. Biol.
4:450-455

Sanchez, I.E., and T. Kiefhaber. 2003. Evidence for sequential barriers and
obligatory intermediates in apparent two-state protein folding. J. Mol.
Biol. 325:367-376

Shastry, M.C.R., and H. Roder. 1998. Evidence for barrier-limited protein
folding kinetics on the microsecond time scale. Nat. Struct. Biol. 5: 385-392

Wagner, C., and T. Kiefhaber. 1999. Intermediates can accelerate protein
folding. Proc. Natl. Acad. Sci. USA 96: 6716-6721

Winter, R.B., O.G. Berg, and P.H. von Hippel. 1981. Diffusion-driven
mechanisms of protein translocation on nucleic acids. III. The E. coli lac
repressor-operator interaction: kinetic measurements and conclusions.
Biochemistry, 20:6961-6977

Wright, C.F., K. Lindorff-Larsen, L.G. Randles, and J. Clarke. 2003.
Parallel protein-unfolding pathways revealed and mapped. Nat. Struct. Biol.
10:658-662

Zhou, Y., and M. Karplus. 1999. Interpreting the folding kinetics of helical
proteins. Nature 401:400-403

Zwanzig, R., 1995. Simple model of protein folding kinetics. Proc. Natl.
Acad. Sci. USA, 21;9801-9804

\begin{acknowledgement}
I would like to thank Sheena Radford, Kevin Plaxco and Joan Shea for helpful
discussions and comments on the manuscripts, and the Kavli Institute for
Theoretical Physics for the hospitality of its 2002 programme ``Dynamics of
Complex and Macromolecular Fluids'', where the bulk of this work was done
(preprint number NSF-ITP-02-55).
\end{acknowledgement}

\pagebreak \pagebreak

\bigskip {\Large Figure Legend}

Figure 1

Part of the $d$-dimensional folding space containing a diffusive hypergutter
projected onto $2$ dimensions. \ The diffusing particle (representing the
random search of the protein through its available conformations) does not
have to search simulataneously through both the dimensions of the figure. \
Instead, it \ exploits the lower energy state of the entire diffusive
subspace of the $(d-n-1)^{th}$ subspace to reach it \textit{via} a
one-dimensional diffusion in the $(d-n)^{th}$ dimension.

\bigskip

Figure 2

The 3-helix bundle (BdpA on the left) is coarse-grained to a system of three
rods. \ The three angles constituting the diffusive subspaces are labelled $%
X_{i}$ for $i=1,2,3$. \ The folding space then looks like the periodic cubic
lattice on the right (only the $X_{2}$ direction is shown periodic, for
clarity). \ The attractive gutter is the 2-d space spanned by $X_{1}$ and $%
X_{3}$ once $X_{2}$-diffusion has brought the third helix into contact with
the other two. \ But for small angles $X_{2}$ there is a large topological
barrier between the ``correct'' and ``incorrect'' sides of attachment of the
third helix onto the bundle formed by the other two, and identical with the
rapid diffusional subspace of $X_{1}$ and $X_{3}$.

\bigskip

Figure 3

Log-linear plot of three relaxation functions. \ Dashed is the single
exponential. \ Dotted is the decay to the native state for 1-dimensional
diffusion with a uniform distribution of initial states. \ Solid curve is
the decay of an effective 1-dimensional folding path created from a high
dimensional landscape with flat free energy.

\bigskip

Figure 4

The one-dimensional folding free-energy in terms of sequential diffusive
subspaces. (a) $T<T_{f}$; the landscape is a directed random walk - the
maximum excursion (transition state) lies towards the start of the walk. (b) 
$T=T_{f}$; the walk returns to the origin of free energy - the maximum lies
near the middle of the trajectory.

\bigskip

Figure 5

Dimensionless Eyring plot of the universal form in the version of the
gutter-landscape model in which only the energies of native and denatured
states are specified (to convert to numbers on an experimental Arrhenius
plot, the figures on the ordinate should be multplied by the square root of
the number of diffusive dimensions, or contact order, of the protein). \ The
inset shows experimental results on the protein CI2, from \cite{Fersht}.

\bigskip

Figure 6

Predictions of the folding rate (solid line) relative to the rate of the
optimal 1+2 dimensional search path, and sum of non-natively generated $\Phi 
$-values from residues contributing to a 2-d diffusive hypergutter (3 helix
bundle) (dashed line). \ The ordinate is the ``fugacity'' measure of the
attractive potential $\Delta =\sigma e^{\varepsilon kT}$. \ The value
assumed for the spatial reduction $\sigma $, is $0.1$. The inset contains
the expected magnitude of increase in $\Phi $-values with residue index
(dashed lines) in a 3-helix protein with a 2-d kinetic intermediate gutter
(Im7), relative to one without (Im9). \ We expect the modifications to be
concentrated onto helices 1 and 4, whose mutual contacts stabilise the 2D
search space.

\clearpage

\begin{figure}[tbp]
\begin{center}
\resizebox{5.0in}{!}{
\includegraphics{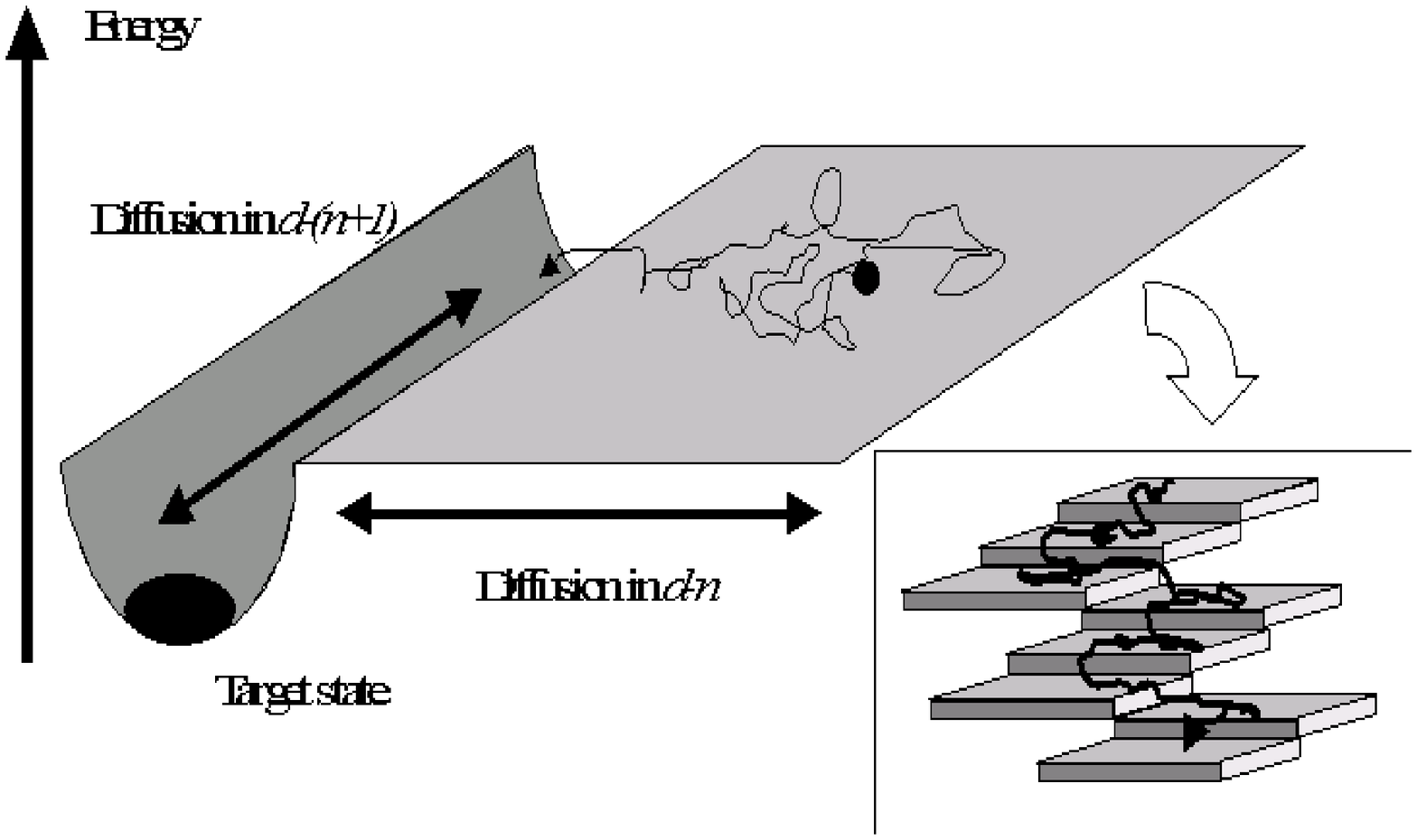}
}
\caption{Schematic of sequential searches in gutters}
\end{center}
\end{figure}

\clearpage

\begin{figure}[tbp]
\begin{center}
\resizebox{5.0in}{!}{
\includegraphics{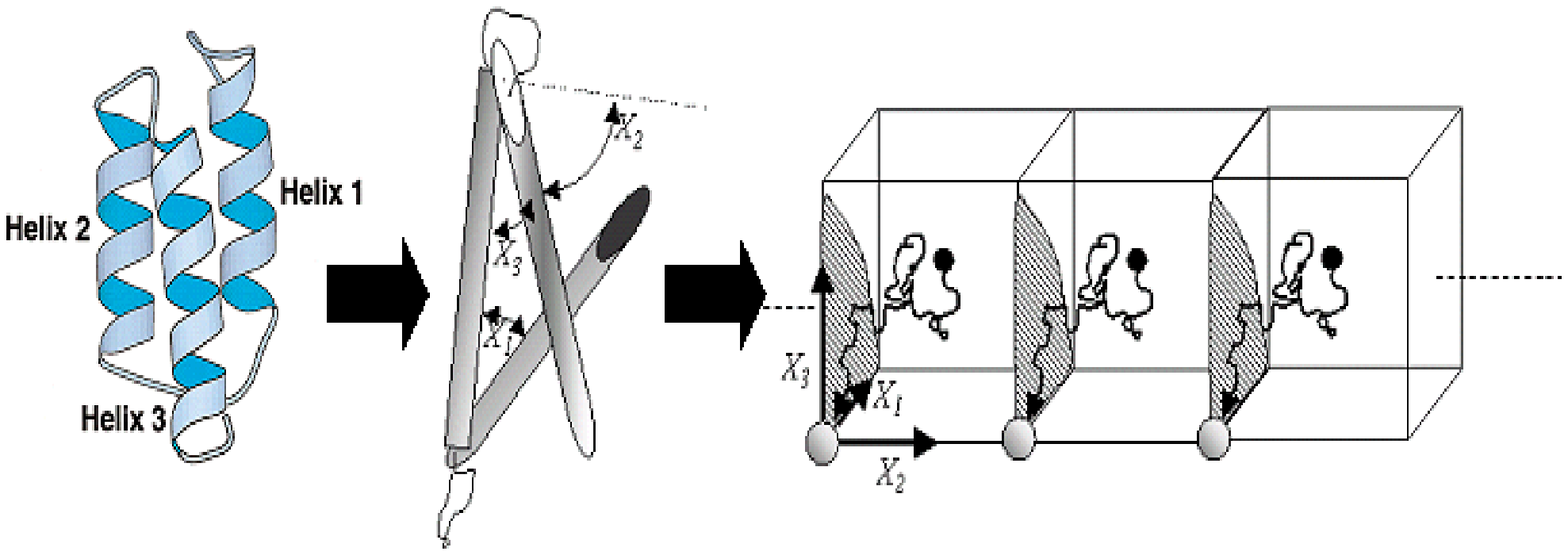}
}
\caption{Coarse-graining of the 3-helix bundle}
\end{center}
\end{figure}

\clearpage

\begin{figure}[tbp]
\begin{center}
\resizebox{4.0in}{!}{
\includegraphics{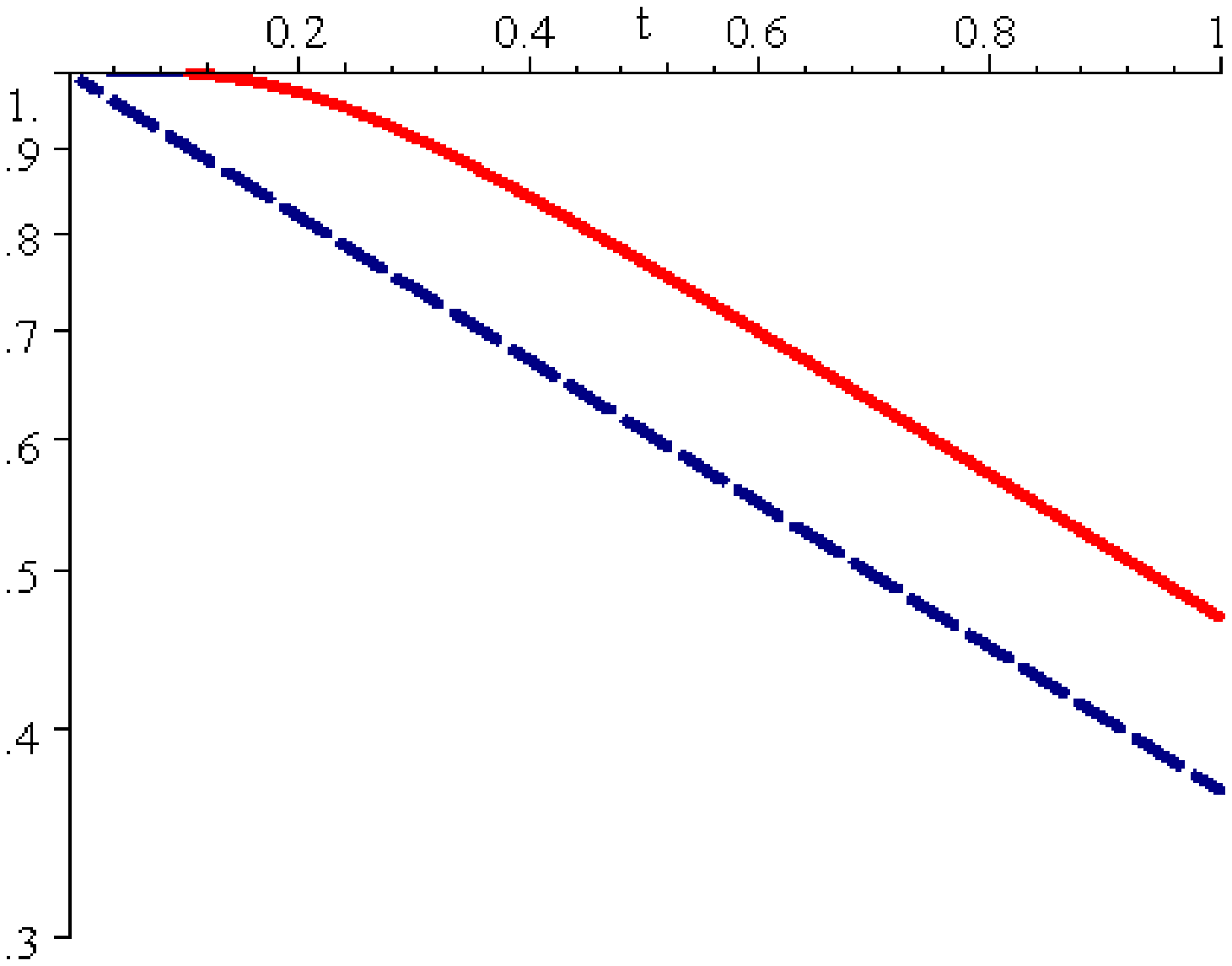}
}
\caption{Delay-correction to single-exponential kinetics}
\end{center}
\end{figure}

\clearpage

\begin{figure}[tbp]
\begin{center}
\resizebox{5.0in}{!}{
\includegraphics{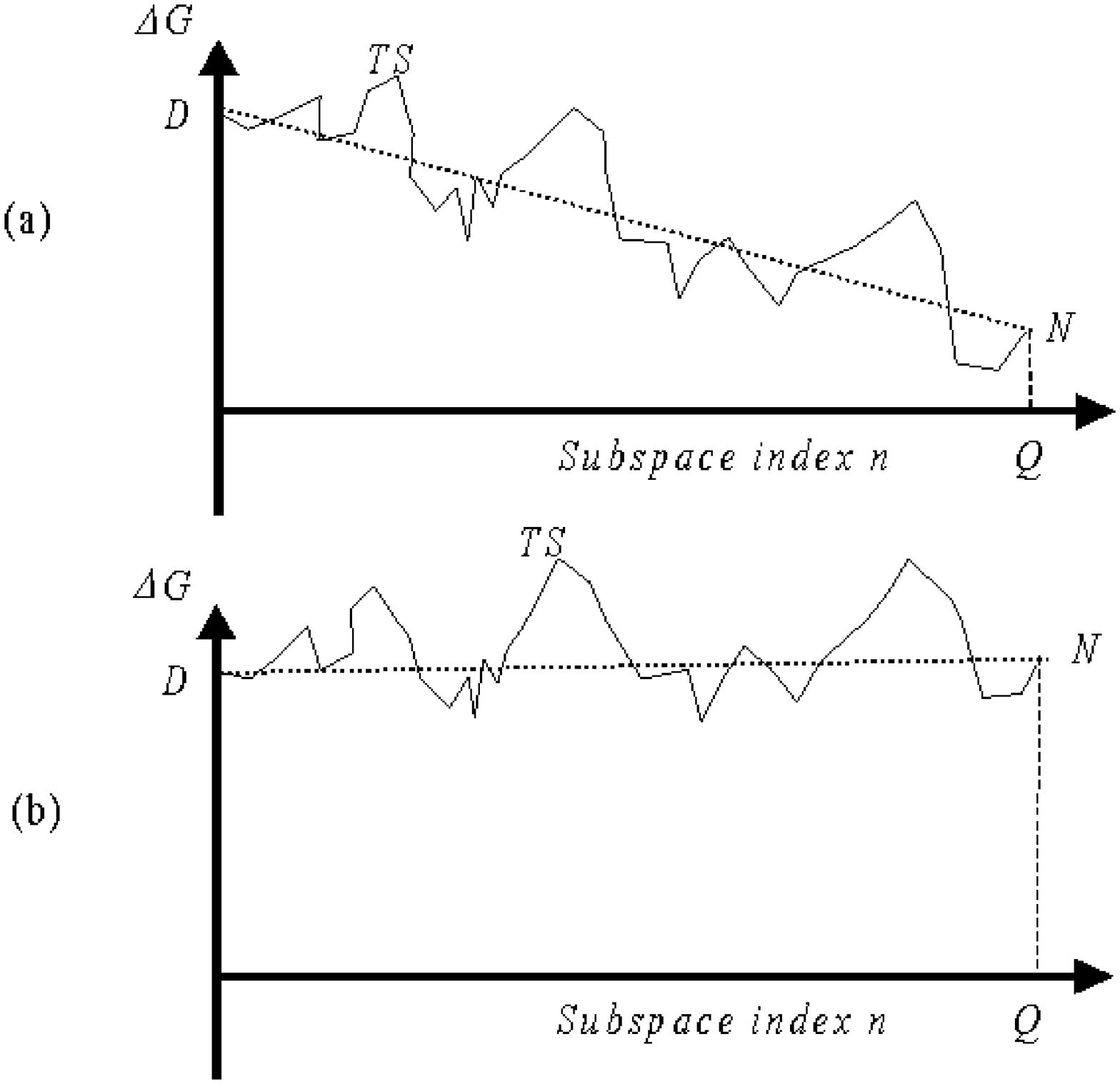}
}
\caption{Random walk of gutter energy landscape}
\end{center}
\end{figure}

\clearpage

\begin{figure}[tbp]
\begin{center}
\resizebox{4.0in}{!}{
\includegraphics{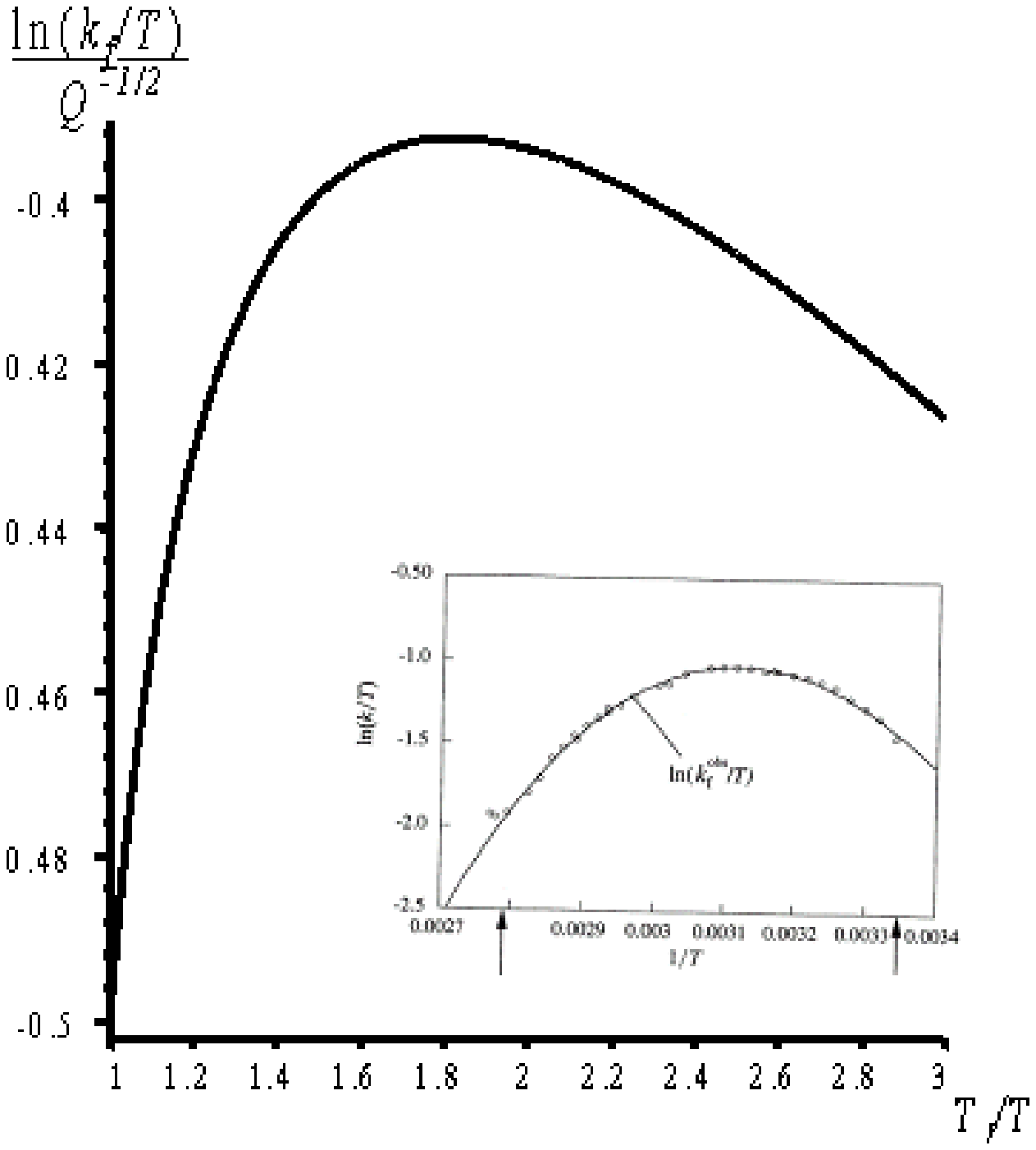}
}
\caption{Predicted curvature of Eyering plot}
\end{center}
\end{figure}

\clearpage

\begin{figure}[tbp]
\begin{center}
\resizebox{5.0in}{!}{
\includegraphics{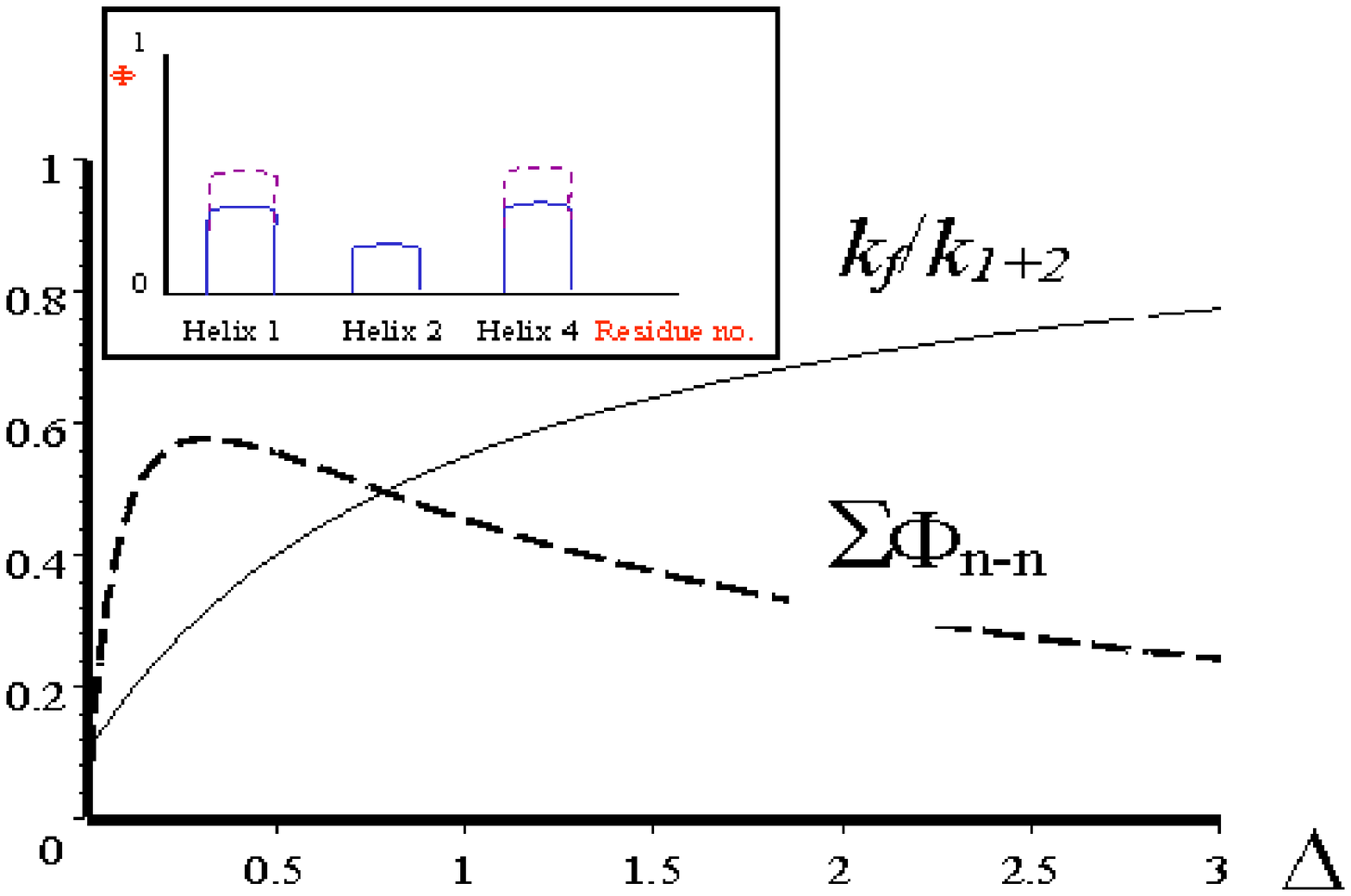}
}
\caption{Dependence of folding rate and effective phi-value on gutter
potential}
\end{center}
\end{figure}

\end{document}